\documentclass[prb,twocolumn,showpacs,amssymb,superscriptaddress]{revtex4-2}
\usepackage{graphicx}
\usepackage{amsmath}
\usepackage{amsfonts}
\usepackage{amsthm}
\usepackage{amssymb}
\usepackage{amsbsy}
\usepackage{wasysym}
\usepackage{bm}
\usepackage{bbm}
\usepackage{mathrsfs}
\usepackage{color}
\usepackage{hyperref}
\usepackage{braket}
\setcitestyle{super}

\newtheorem{defn}{Definition}
\newtheorem{theorem}{Theorem}
\newtheorem{prop}{Proposition}

\newcommand*{\citen}[1]{%
  \begingroup
    \romannumeral-`\x 
    \setcitestyle{numbers}%
    \cite{#1}%
  \endgroup   
}

\date{\today}
\begin{document}

\title{How smooth is quantum complexity?}

\author{Vir B. Bulchandani}
\affiliation{Princeton Center for Theoretical Science,  Princeton University, Princeton, New Jersey 08544, USA}

\author{S. L. Sondhi}
\affiliation{Department of Physics, Princeton University, Princeton, New Jersey 08544, USA}

\begin{abstract}
The ``quantum complexity'' of a unitary operator measures the difficulty of its construction from a set of elementary quantum gates. While the notion of quantum complexity was first introduced as a quantum generalization of the classical computational complexity, it has since been argued to hold a fundamental significance in its own right, as a physical quantity analogous to the thermodynamic entropy. In this paper, we present a unified perspective on various notions of quantum complexity, viewed as functions on the space of unitary operators. One striking feature of these functions is that they can exhibit non-smooth and even fractal behaviour. We use ideas from Diophantine approximation theory and sub-Riemannian geometry to rigorously quantify this lack of smoothness. Implications for the physical meaning of quantum complexity are discussed.
\end{abstract}

\maketitle

\section{Introduction}

The computational complexity of a classical algorithm measures the physical resources that the algorithm requires to run. Although its precise definition depends on the model of computation in question, the classical computational complexity is robust in the sense that equivalent models of computation give rise to equivalent complexity measures, up to an overhead in computational resources that is polynomial in the input size. For example, the computational complexity of an algorithm on $N$ input bits can be defined as its halting time $C_1(N)$ on a Turing machine or as the number of elementary Boolean operations $C_2(N)$ in its Boolean circuit representation. Equivalence of the Turing machine and Boolean circuit models of classical computation then guarantees that $C_2(N) = \mathcal{O}(\mathrm{poly}(C_1(N))$.
This idea of equivalence up to polynomial functions yields a notion of ``asymptotic complexity'' that is independent of one's preferred model of classical computation.

The extension of these classical definitions to quantum computers is not unique, allowing for various notions of quantum complexity with very different properties. A quantum algorithm on $N$ qubits can be thought of as a unitary operator on the $N$-qubit Hilbert space; in this sense, each definition of quantum complexity yields a distinct function on the space of unitary operators $G = SU(2^N)$. A standard choice, that mimics the classical circuit complexity, is to fix some set of elementary gates $\mathcal{A} \subset G$ and define the complexity $C(U)$ of an operator $U \in G$ as the number of elementary gates that must be multiplied to obtain $U$. However, this formulation of quantum complexity is rife with ambiguities that do not arise in the classical setting. For example, the set of elementary gates $\mathcal{A}$ can be chosen to be discrete and finite or continuous and infinite, leading to very different functions $C(U)$ in each case. Nevertheless, this notion of quantum computational complexity turns out to be robust enough for practical purposes, allowing one to define quantum complexity classes by analogy with their classical counterparts~\cite{nielsenchuang}.

Thus, at least from the viewpoint of quantum computer science, such ambiguities in the definition of quantum complexity $C(U)$ are not very serious. However, recent work by Susskind and collaborators on the late-time dynamics of black holes~\cite{SusskindBlackHoles16,secondlaw,susskind2018lectures,susskind2020black} has argued that the quantum complexity should be viewed as a fundamental physical quantity, analogous to the usual thermodynamic entropy in some respects but with important differences in others~\cite{secondlaw}. In particular, it has been proposed\cite{ComplexityEqualsAction} that within the AdS/CFT correspondence, the dynamics of quantum complexity on the boundary of AdS spacetime uniquely captures certain non-trivial late-time features of black hole dynamics in the bulk, long after the timescale at which thermalization sets in.

If correct, this intriguing conjecture has consequences far beyond the specific context of black hole physics. First, it implies that there may be a universal definition of quantum complexity that is determined by physical principles alone. This stands in contrast to the usual definition of classical computational complexity, which is rather arbitrary from the physicist's viewpoint. Second, it suggests that the quantum complexity might be a useful tool in the broader setting of many-body physics, for example, in distinguishing highly entangled states of condensed matter systems. Both these possibilities call for a study of the quantum complexity as a physically interesting quantity in its own right.

With this motivation, our paper provides a unified perspective on the various notions of quantum complexity. We first show how three standard measures of the complexity of a quantum circuit -- the discrete gate complexity, the circuit depth and the continuous complexity distance -- define related generalizations of classical computational complexity to the quantum setting. We next explore the analytical properties of these complexity measures, viewed as functions on the space of unitary operators, which have previously been argued to exhibit non-smooth and possibly fractal behaviour~\cite{susskind2018lectures,singlequbit,susskind2020black}. The main technical result of our paper is a rigorous account of this lack of smoothness. 


\section{Three notions of quantum complexity}

The classical notion of computational complexity, for example based on Boolean circuits, operates in discrete time with a discrete set of elementary operations. In contrast, the unitary group has the structure of a smooth manifold. Thus in attempting to define the computational complexity of a quantum algorithm on $N$ qubits, viewed as a unitary operator $U \in G=SU(2^N)$, there is no \emph{a priori} reason to restrict oneself to either a discrete notion of computational ``time'' or a discrete set of elementary operations, beyond a desire for harmony with established classical definitions.

Successively relaxing these requirements of discreteness in time and discreteness of gates leads naturally to three distinct notions of quantum complexity. First let us recall that there are two fundamentally distinct types of gate set $\mathcal{A} \subset G$: 
\begin{enumerate}
    \item \emph{exactly universal}: $\mathcal{A}$ is infinite and uncountable, and every $U \in G$ can be expressed as a finite product of elements in $\mathcal {A}$.
    \item \emph{computationally universal}: $\mathcal{A}$ is finite, and the group generated by elements of $\mathcal{A}$ and their inverses is dense in $G$.
\end{enumerate}

One further distinction, of potential importance for applications of quantum complexity to condensed matter systems or local quantum fields, is between spatially local and $k$-local gate sets. Previous discussions of the quantum complexity tend to focus on elementary gate sets $\mathcal{A}$ that are $k$-local and all-to-all\cite{nielsen2005geometric,susskind2018lectures}. Demanding instead that the gate set $\mathcal{A}$ be local in space leads to some small differences compared to the non-local case, which we address in later sections.

The distinction between computationally universal and exactly universal gates sets leads to correspondingly distinct notions of quantum complexity:
\begin{figure*}[t]
    \centering
    \includegraphics[width=0.99\linewidth]{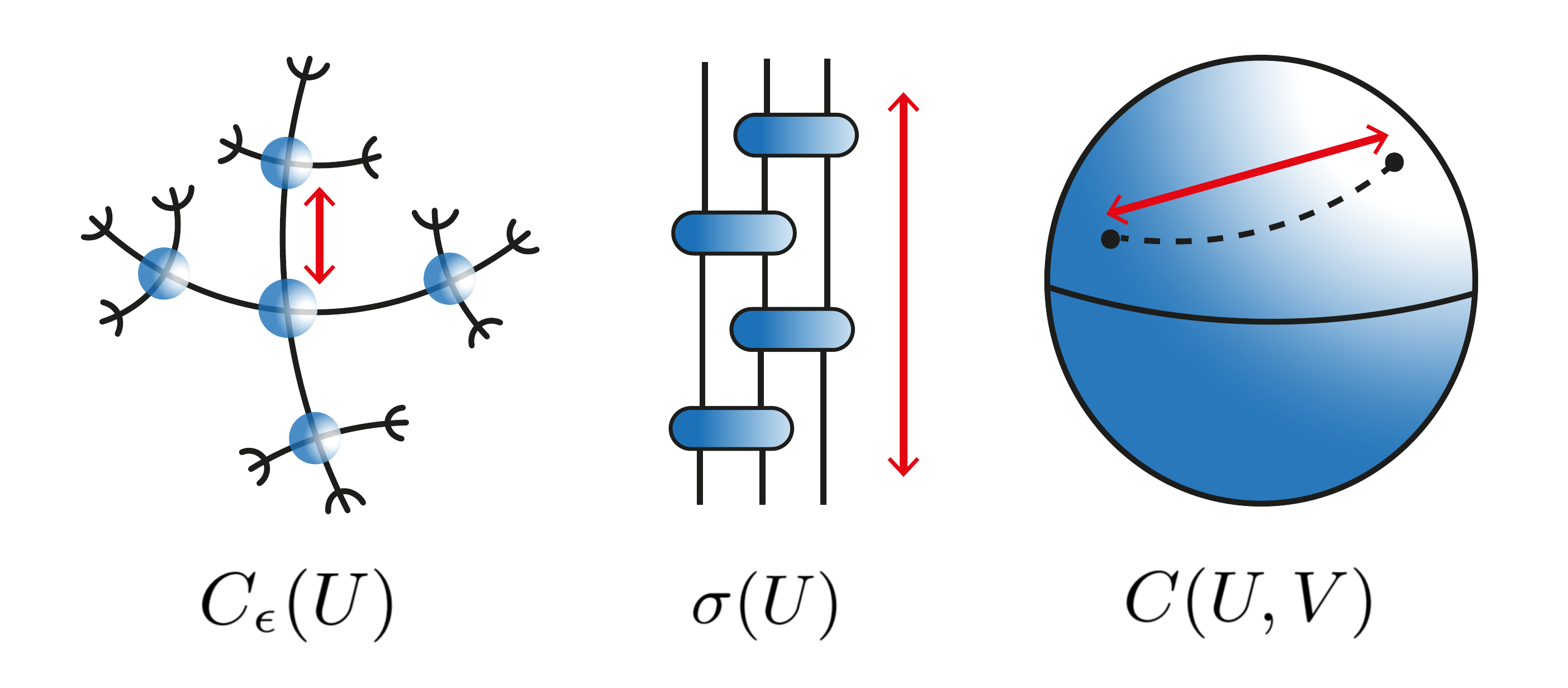}
    \caption{Left to right: the gate complexity, the circuit depth and the complexity distance define three increasingly smooth notions of quantum complexity on the space of unitary operators.}
    \label{Fig1}
\end{figure*}

\begin{enumerate}
    \item (discrete time, discrete gates) Let the gate set $\mathcal{A} = \{g_1,g_2,\ldots,g_r\} \subset G$ be computationally universal. Then for a given tolerance $\epsilon>0$, the \emph{gate complexity} of a unitary operator $U \in G$ is defined as
\begin{equation}
\label{eq:gate_complexity}
    \mathcal{C}_{\epsilon}(U) = \mathrm{min}\{\sum_{j=1}^k |n_j| : \| g_{i_1}^{n_1}g_{i_2}^{n_2}\ldots g_{i_k}^{n_k}-U\| < \epsilon \}.
\end{equation}
Here the minimization is over all $k\in \mathbb{N}$,  $i_j \in \{1,2,\ldots,r\}$ and $n_1,n_2,\ldots,n_k \in \mathbb{Z}$. The matrix norm can be chosen freely; in this paper we will use the Hilbert-Schmidt norm $\|A\| = [\mathrm{Tr}(A^\dagger A)]^{1/2}$. Among the three definitions of quantum complexity considered here, the gate complexity is most closely analogous to the classical computational complexity.

\item (discrete time, continuous gates) Now suppose that the gate set $\mathcal{A}$ is exactly universal, but computational ``time'' is discrete, and at most $W = \mathcal{O}(\mathrm{poly}(N))$ gates may be applied in a given time step. For example, $\mathcal{A}$ might be the set of all operators on two qubits, of which at most $W=N/2$ can be applied simultaneously. In this setting, a natural measure of the quantum complexity is the \emph{circuit depth} $\sigma(U)$ of the shallowest circuit required to simulate $U$. Up to a factor polynomial in $N$, this is equivalent to the least number of distinct gates in $\mathcal{A}$ required to simulate $U$.

\item (continuous time, continuous gates) Finally, one can consider both an exactly universal gate set $\mathcal{A}$ and allow the unitary $U$ of interest to be simulated in continuous time. The ``cost'' of simulating $U$ is then most naturally formulated as a problem in Hamiltonian control\cite{Khaneja,nielsen2005geometric}. Concretely, one introduces a cost function $f(H)$ that defines a norm on the space of $2^N \times 2^N$ Hermitian matrices, and for given $U, V \in G$, defines their \emph{complexity distance} $C(U,V)$ in terms of the least costly path from the identity $U$ to $V$ in $G$, to wit
\begin{equation}
C(U,V) = \inf_{\gamma \, \mathrm{a.c.}} \int_0^1 dt \, f(H),
\end{equation}
where the infimum is over all absolutely continuous~\cite{ledonne2010lecture} paths $\gamma: [0,1] \to G$ such that
\begin{equation}
\frac{d}{dt} \gamma = -iH(t) \gamma, \quad \gamma(0) = U, \, \gamma(1) = V.
\end{equation}
The complexity of a specific operator $U \in G$ is then given by $C(U) := C(\mathbbm{1},U)$, where $\mathbbm{1}$ denotes the identity in $G$. This formulation of quantum complexity is usually referred to as ``complexity geometry''. For suitable choices of $f$, $C(U,V)$ is the geodesic distance between $U$ and $V$ with respect to a Riemannian metric on $G$. More generally, $f$ endows $G$ with the structure of a sub-Finsler manifold, with respect which the complexity distance $C(U,V)$ defines a so-called Carnot-Carath{\'e}odory distance~\cite{Gromov1996,ledonne2010lecture}.
\end{enumerate}

These three broad notions of quantum complexity and their basic properties are summarized in Fig. \ref{Fig1} and Table \ref{Tab1}. Among these definitions, viewed as functions on $G$, the circuit depth is simplest to capture analytically\cite{Lloyd}. A particularly tractable example is the case of a single qubit $G=SU(2)$, for which the gate set $\mathcal{A} = \{e^{i\phi Z},e^{i\theta Y} : \phi,\,\theta \in [0,2\pi)\}$ is exactly universal by Euler angles. Then unitaries $U \in G$ that are generic in measure are realized by circuits of minimum depth $\sigma(U)=3$, corresponding to their Euler angle parameterization $U = e^{i\chi_1 Z} e^{i\chi_2 Y} e^{i\chi_3 Z}$. More generally, consider $N$ qubits with $G = SU(2^N)$ and suppose that the exactly universal gate set $\mathcal{A}$ in question is parameterized by $K = \mathcal{O}(\mathrm{poly}(N))$ continuous parameters. Since the real dimension of the manifold $G$ is $4^N-1$, the circuit depth of a unitary $U$ that is generic in measure must satisfy the inequality
\begin{equation}
\sigma(U) \geq \frac{\dim{G}}{K W} \sim \frac{4^N}{\mathrm{poly}(N)}, \quad N \to \infty.
\end{equation}
Thus the circuit depth $\sigma(U)$ is bounded below almost everywhere by a function exponentially large in $N$. 

In fact, both the discrete gate complexity $C_{\epsilon}(U)$ and the continuous complexity distance $C(U,V)$ exhibit similar worst-case exponential scaling~\cite{susskind2018lectures} in $N$ to the circuit depth $\sigma(U)$. It is therefore usually assumed\footnote{More precisely, for an exactly universal gate set consisting of all one and two qubit gates, it is known that provided the penalty factor $q$ in Nielsen's metric is taken sufficiently large~\cite{QuantCompasGeom,dowling2008geometry}, e.g. $q>4^N$, the estimate $C(U) = \mathcal{O}(\mathrm{poly}(N)\sigma(U))$ holds. The gate complexity defined with respect to an efficiently universal gate set satisfies a similar estimate $C_{\epsilon}(U) = \mathcal{O}(\log{(1/\epsilon)} \sigma(U))$. Equivalence between the gate complexity and the complexity distance at all scales is unlikely for the reasons discussed in Section \ref{sec:obstacle}.} that these definitions are more-or-less equivalent from the viewpoint of quantum computing~\cite{nielsenchuang,QuantCompasGeom}, and differ only in their ease of calculation.
However, if the complexity is viewed as a fundamental physical quantity in its own right~\cite{ComplexityEqualsAction}, then the distinction between these different complexity measures becomes important. 

The most striking such distinction arises in the local analyticity properties of these complexity measures. In particular, both the gate complexity and the complexity distance have a more intricate mathematical structure than the circuit depth and can exhibit highly non-smooth behaviour. The purpose of the remaining sections is to quantify this lack of smoothness rigorously. Our results are organized as follows. We first prove the existence of efficiently universal gate sets for which $C_{\epsilon}(U)$ exhibits worst-case scaling in $\epsilon$, $C_{\epsilon}(U) = \Omega(\log{1/\epsilon})$, densely in $G$. This clarifies the manner in which $C_{\epsilon}(U)$ tends to a nowhere continuous function as $\epsilon \to 0$. We next prove that the complexity distance $C(U,V)$ is at worst continuous but not differentiable as $V \to U$ in generic directions. These results give precise meaning to various qualitative discussions of the ``fractal'' geometry of quantum complexity in the literature.

\begin{table*}[t]
\begin{tabular}{|c|c|c|c|c|}
\hline
    Time & Gate set & Complexity measure & Notation & Mathematical interpretation \\
    \hline
     Discrete & Discrete & Gate complexity & $C_{\epsilon}(U)$ & Regularized word metric \\
     Discrete & Continuous & Circuit depth & $\sigma(U)$ & Dimension of group \\
     Continuous & Continuous & Complexity distance & $C(U,V)$ & Carnot-Carath{\'e}odory distance\\
     \hline
\end{tabular}
\caption{A summary of the various notions of quantum complexity that we consider in this paper.}
\label{Tab1}
\end{table*}

\section{Gate complexity}
\subsection{Background to results}

For the gate complexity Eq. \eqref{eq:gate_complexity}, let $\langle \mathcal{A} \rangle$ denote the group generated by elements of $\mathcal{A}$ and their inverses, which by assumption is dense in $G$. To gain some intuition, first consider the limit $C_0(U) := \lim_{\epsilon \to 0} C_\epsilon(U)$. Because the group $\langle \mathcal{A} \rangle$ has measure zero in $G$, $C_0(U) = \infty$ almost everywhere. Thus as $\epsilon \to 0$, $C_\epsilon(U)$ tends to a nowhere continuous function on the group $G$, analogous to the nowhere continuous Dirichlet function of real analysis. (Notice that for a finite group with $\langle A \rangle = G$, $C_0$ recovers the ``word metric'' studied in geometric group theory; we refer to Ref. \citen{Lin2019} for a discussion of analogies between geometric group theory and quantum complexity.)

The lack of continuity of $C_0$, viewed as a function from $G$ to the extended real line, is usually assumed to be regulated by the introduction of $\epsilon>0$; in this sense the quantum complexity defines a regularized notion of word metric for Lie groups. Indeed, for any computationally universal gate set, the Solovay-Kitaev theorem provides a constructive algorithm~\cite{SolKit} to approximate any given $U \in G$ within $\epsilon>0$ with $C_\epsilon(U) = \mathcal{O}(\log^{c}{1/\epsilon})$ where $c \approx 4$. It was shown by Harrow, Recht and Chuang~\cite{Harrow} that the Solovay-Kitaev estimate could be improved for the class of so-called ``efficiently universal'' gate sets, which allow all unitaries $U$ to be $\epsilon$-approximated using at most $C_\epsilon(U) = \mathcal{O}(\log{1/\epsilon})$ elementary gates. This result was also shown to be optimal, in the sense that the worst-case complexity of $U \in G$ is bounded below by $\Omega(\log{1/\epsilon})$.

The $\log{1/\epsilon}$ scaling of the worst-case complexity follows by an elementary counting argument, which works by thickening points of $\langle A \rangle$ in $G$ by $\epsilon$-balls until the entire group is covered\cite{Harrow}. However, it is not clear from this argument how such ``complicated unitaries'' are distributed within the Lie group $G$. The main result of this section is that there exist efficiently universal gate sets for which complicated unitaries are dense in $G$. In other words, every unitary in $G$ is arbitrarily close to another unitary whose complexity scales as $\Omega(\log{1/\epsilon})$ as $\epsilon\to 0$. Since $\langle A \rangle$ is also dense in $G$, it follows that every neighbourhood of $G$ contains both a unitary with finite complexity independent of $\epsilon$ and a unitary with complexity diverging as $\log{1/\epsilon}$ as $\epsilon \to 0$. This illustrates how the discontinuous behaviour of $C_0$ is inherited by the regularized gate complexity, $C_{\epsilon}$ with $\epsilon>0$.

In fact, the mathematical idea necessary to capture this behaviour did not appear until relatively recently. The foundation of our analysis is the ``non-commutative Diophantine property'' introduced by Gamburd, Jakobson and Sarnak\cite{Sarnak} and subsequently refined by Bourgain and Gamburd\cite{Bourgain2008,bourgain2012spectral}, that is essentially a non-Abelian analogue of the Diophantine properties satisfied by the real numbers\cite{schmidt2009diophantine}. As a rule, Diophantine properties tend to be the preserve of number theorists, rather than physicists; a notable exception is the KAM theory of perturbed integrable systems, which demonstrates that quasiperiodic tori whose frequencies satisfy Diophantine bounds are stable to the onset of chaos~\cite{poschel}. In the complexity language, KAM tori have ``complicated resonances", that cannot be achieved at any finite order in perturbation theory. Other examples include the sensitivity of the Aubry-Andr{\'e} model to the commensuration of its potential~\cite{aubry1980analyticity} and the sensitivity of the quasiparticle content of the gapless spin-$1/2$ XXZ chain to its anisotropy~\cite{takahashi_book}. Our analysis of the gate complexity $C_{\epsilon}(U)$ exploits its similarity to these phenomena, insofar as the gate complexity is also sensitive to the irrationality properties of the real numbers\cite{Sarnak,Harrow,Bourgain2008,bourgain2012spectral,GoldenGates}.

For the special case $G = U(1)$, lower bounds on the gate complexity follow by standard results in the Diophantine approximation of real numbers\cite{schmidt2009diophantine}. The extension to gate complexities in non-Abelian unitary groups, $G = SU(d)$ with $d\geq 2$, is only made possible by some relatively recent mathematical advances in Lie group theory\cite{Sarnak,Breuillard,Bourgain2008,bourgain2012spectral}, that were alluded to above. The remainder of this section is structured as follows. We first familiarize the reader with Diophantine bounds in the simplified setting of $U(1)$, which contains most of the ideas necessary for understanding the non-Abelian case. We then provide a worked example of the singular behaviour of $C_{\epsilon}(U)$ for the basic physical case of $G = SU(2)$. Finally, we extend the result to a class of efficiently universal gate sets in $G=SU(d)$ with $d>2$. 

\subsection{Gate complexity in $U(1)$}
The basic ideas of our proof in the non-Abelian case can be understood by applying standard results in Diophantine approximation\cite{schmidt2009diophantine} to lower bound the gate complexity on a dense subset of the group $G=U(1)$. Due to the projective nature of quantum mechanics, this example is somewhat unphysical, but is nevertheless instructive. 

When $G = U(1)$, a single generator $g_1 = e^{i\alpha}$ with $\alpha/\pi$ irrational is computationally universal, in the sense that $\langle g_1 \rangle$ is dense in $U(1)$. The complexity of $U = e^{i\phi} \in G$ associated with the gate set $\mathcal{A} = \{g_1\}$ is given by
\begin{equation}
C_{\epsilon}(e^{i\phi}) = \min\{|n| : |e^{i\phi} - e^{in\alpha}| < \epsilon\}.
\end{equation}
Thus we consider $n$ such that
\begin{equation}
\left| \sin \left(\frac{\phi - n \alpha}{2}\right) \right| < \frac{\epsilon}{2}.
\end{equation}
Let $m$ denote the unique integer such that
\begin{align}
\frac{\phi - n\alpha}{2\pi} - \frac{1}{2} \leq m < \frac{\phi - n \alpha}{2\pi} + \frac{1}{2}.
\end{align}
Then
\begin{equation}
\left| \frac{\phi - n\alpha}{2} - m\pi\right| < \frac{\pi}{2}.
\end{equation}
Using the inequality $(2/\pi)|x| < |\sin x|$ for $|x| < \pi/2$, it follows that
\begin{align}
\label{eq:ub}
\left| \frac{\phi}{2\pi} - n \frac{\alpha}{2\pi} - m \right| < \frac{\epsilon}{4}.
\end{align}
To proceed further, we make the additional assumption that $\frac{\alpha}{2\pi}$ is an algebraic irrational number. We then have the following result:

\begin{prop} Let $\phi \in [0,2\pi)$ and $\delta >0$. Then there exists $\phi'$ with $|\phi'-\phi|<\delta$ such that 
\begin{equation}
\log C_{\epsilon}(\phi') > \frac{1}{3} \log{\left(\frac{1}{\epsilon}\right)} + \mathcal{O}(\epsilon^0).
\end{equation}
\end{prop}

\textit{Proof.} Since algebraic numbers are dense in $\mathbb{R}$ and $\alpha/2\pi$ has finite degree by assumption, there exists $\phi'$ algebraic such that $|\phi' - \phi|< \delta$ and $\phi'/2\pi$ is rationally independent from $\alpha/2\pi$. Let $n \in \mathbb{Z}$ with $|e^{i\phi'} - e^{in\alpha}| < \epsilon$. Then $\phi'$ satisfies Eq. \eqref{eq:ub} for some $m \in \mathbb{Z}$. Also, by a standard result in Diophantine approximation\cite{schmidt2009diophantine}, for any exponent $\tau>2$ there is a constant $K(\alpha,\phi',\tau)$ such that
\begin{equation}
\left| \frac{\phi'}{2\pi} - n \frac{\alpha}{2\pi} - m \right| > \frac{K(\alpha,\phi',\tau)}{(|n|+|m|+1)^\tau}.
\end{equation}
Using the definition of $m$, it follows by Eq. \eqref{eq:ub} that
\begin{equation}
\label{eq:dioph}
\log C_{\epsilon}(\phi') > \frac{1}{\tau} \log{\left(\frac{1}{\epsilon}\right)} + \mathcal{O}(\epsilon^0).
\end{equation}
The result follows upon setting $\tau=3$. $\square$

\subsection{Gate complexity in $SU(2)$}
Let us now turn to the more difficult case of $G=SU(2)$. For concreteness, consider the gate set $\mathcal{A}_2 = \{e^{i\pi \alpha Z}, e^{i\pi \alpha Y}\}$ where $Y,Z$ denote the usual Pauli matrices and $\cos{\pi \alpha} \in \mathbb{Q} \backslash \{0,\pm 1,\pm 1/2 \}$. It follows by Niven's theorem that $\alpha$ is irrational; thus the gates in $\mathcal{A}$ generate dense subsets of the circles $\{e^{itZ}, t \in [0,2\pi)\}$ and $\{e^{itY}, t \in [0,2\pi)\}$. By Euler angles, $\langle \mathcal{A}_2 \rangle$ is dense in $G$ and $\mathcal{A}_2$ yields a computationally universal gate set on $SU(2)$.

Moreover, by a theorem of Swierczkowski\cite{Swierczkowski}, the group $\langle \mathcal{A}_2 \rangle$ is \emph{free}; this means that its Cayley graph is a tree and therefore defines a (rooted) Bethe lattice with coordination number $z=4$ in the compact manifold $SU(2)$. This ``fractal'' geometry suggests discontinuous behaviour of the quantum complexity, as has been suggested in the literature~\cite{susskind2018lectures}; we now make this intuition precise.

The main technical apparatus that we will require is as follows:

\begin{defn}\emph{(Gamburd, Jakobson, Sarnak\cite{Sarnak})}
Let $\mathcal{A} = \{g_1,g_2,\ldots, g_r\}$ with $g_i \in SU(2)$. $\mathcal{A}$ satisfies the \emph{non-Abelian Diophantine condition} if there exists a constant $D = D(\mathcal{A})$ such that for all words $W_l = g_{i_1}g_{i_2}\ldots g_{i_l} \neq \pm \mathbbm{1}_2$ of length $l$ in $\langle \mathcal{A} \rangle$, the inequality
\begin{align}
\label{eq:nonAbdiop}
\| W_l \pm \mathbbm{1}_2 \| \geq D^{-l}
\end{align}
holds.
\end{defn}
\begin{prop} \emph{(Gamburd, Jakobson, Sarnak\cite{Sarnak})}
\label{prop:dioph}
Let $\mathcal{A} = \{g_1,g_2,\ldots, g_r\}$ with $g_i \in SU(2) \cap M_2(\overline{\mathbb{Q}})$, where $M_2(\overline{\mathbb{Q}})$ denotes the set of $2 \times 2$ matrices with algebraic elements. Then $\mathcal{A}$ satisfies the non-Abelian Diophantine condition.
\end{prop}
It is clear that the definition Eq. \eqref{eq:nonAbdiop} is analogous to a real Diophantine condition, although the lower bound in the non-Abelian case is exponential rather than polynomial in the word length. It is also clear that our two-element gate set $\mathcal{A}_2 = \{e^{i\pi \alpha Z}, e^{i \pi \alpha Y}\}$ satisfies the non-Abelian Diophantine condition. In fact, this implies that $\mathcal{A}_2$ is efficiently universal\cite{Harrow,Bourgain2008}, in the sense that any element of $SU(2)$ can be $\epsilon$-approximated with worst-case complexity $\mathcal{O}(\log{1/\epsilon})$. We now demonstrate that unitaries with complexity $\Omega(\log{1/\epsilon})$ are dense in $SU(2)$.

\begin{prop} 
\label{prop:su2}
Let $U \in SU(2)$ and $C_\epsilon(U)$ denote its gate complexity defined with respect to $\mathcal{A}_2$. Then for all $\delta > 0$, there exists $U' \in SU(2)$ with $\|U'-U\|<\delta$ and a non-universal constant $D(\alpha,U')$ such that
\begin{equation}
C_{\epsilon}(U') > \frac{1}{\log{D}}\log{\left(\frac{1}{\epsilon}\right)} + \mathcal{O}(\epsilon^0).
\end{equation}
\end{prop}

\textit{Proof}. Pick $U' \in SU(2) \cap M_2(\overline{\mathbb{Q}})$ with $\| U - U'\|<\delta$ and $U' \notin \langle A_2 \rangle$ and let $W^{(2)}_n$ denote a word of length $n$ in $\langle \mathcal{A}_2 \rangle$ that $\epsilon$-approximates $U'$. Then by Proposition \ref{prop:dioph} applied to the gate set $A_2 \cup \{U'\}$, there exists a constant $D(\alpha,U')$ such that
\begin{equation}
\| W_n^{(2)}{U'}^{-1} - \mathbbm{1}_2 \| \geq D^{-(n+1)}.
\end{equation}
By unitarity of $U'$, it follows that
\begin{equation}
\| W_n^{(2)} - U' \| \geq D^{-(n+1)}.
\end{equation}
Then by assumption
\begin{equation}
\epsilon > D^{-(n+1)},
\end{equation}
which implies
\begin{equation}
n+1 > \frac{1}{\log{D}} \log{\left(\frac{1}{\epsilon}\right)}
\end{equation}
(note that it is possible\cite{Sarnak} to set $D>1$). Thus we have achieved a lower bound on the length of words that $\epsilon$-approximate $U'$, and in particular
\begin{equation}
C_{\epsilon}(U') > \frac{1}{\log{D}} \log{\left(\frac{1}{\epsilon}\right)} + \mathcal{O}(\epsilon^0),
\end{equation}
which was to be shown. $\square$

It is immediate that the conclusions of Proposition \ref{prop:su2} extend to any computationally universal gate set $\mathcal{A}=(g_1,g_2,\ldots,g_r)$ on $SU(2)$, provided that the $g_i$ have algebraic elements. As noted previously\cite{Bourgain2008}, all such gate sets are efficiently universal. It follows that our results hold for the entire class of known (at the time of writing) efficiently universal gate sets on $SU(2)$, which includes familiar examples such as the ``Clifford+T'' gate set acting on a single qubit\cite{GoldenGates}.

\subsection{Gate complexity in $SU(d)$}
Finally, we discuss the general (and most physical) case of quantum complexity of unitaries in general special unitary groups, $SU(d)$, with $d>2$.

Unfortunately, this case is also the most difficult and there are not very many explicit results available. We will therefore proceed somewhat indirectly. Let us first note that we can use $\mathcal{A}_2$ defined above for $SU(2)$ to construct an efficiently universal gate set on $SU(d)$, given by the embedding\cite{Harrow}
\begin{equation}
\label{eq:embedding}
\mathcal{A} = \{ \beta_j(M) : j=2,3,\ldots,d,M \in \mathcal{A}_2\},
\end{equation}
where
\begin{equation}
\beta_j(M) = 
\begin{pmatrix}
\mathbbm{1}_{j-2} & 0 & 0 \\
0 & M & 0 \\
0 & 0 & \mathbbm{1}_{d-j}
\end{pmatrix}.
\end{equation}

Since this gate set is efficiently universal, it is certainly computationally universal and therefore generates a dense subgroup of $G = SU(d)$. We then have the following result, which follows by a direct generalization\cite{bourgain2012spectral,breuillard2018expansion} of Proposition \ref{prop:dioph} to $SU(d)$:

\begin{theorem} 
\label{thm:thm1}
Let $\mathcal{A} = \{g_1,g_2,\ldots,g_r\}$ where $g_i \in SU(d) \cap M_d(\overline{\mathbb{Q}})$, with $\langle \mathcal{A} \rangle$ dense in $SU(d)$ and $d \geq 2$. Let $U \in SU(d)$ and $C_\epsilon(U)$ denote its complexity defined with respect to $\mathcal{A}$. Then for all $\delta > 0$, there exists $U' \in SU(d)$ with $\|U-U'\|<\delta$ and a non-universal constant $D(\mathcal{A},U')$ such that
\begin{equation}
\label{eq:sudbound}
C_{\epsilon}(U') > \frac{1}{\log{D}}\log{\left(\frac{1}{\epsilon}\right)} + \mathcal{O}(\epsilon^0).
\end{equation}
\end{theorem}

The proof proceeds similarly to that of Proposition \ref{prop:su2}. Note that the lower bound of Eq. \eqref{eq:sudbound} is ``fractal'', in the sense that the prefactor $D(\mathcal{A},U')$ depends\cite{Sarnak} on the irrationality properties of $\mathcal{A}$ and $U'$. An analogous observation holds for the simpler $U(1)$ case, Eq. \eqref{eq:dioph}, for which the least permissible $\tau$ for given $\phi'$ (not necessarily algebraic) is related to its irrationality measure. Such ``fractal'' lower bounds are quite unusual in physics; a well-studied example arises in the transport theory of the spin-$1/2$ XXZ chain~\cite{Prosen}.

\subsection{Locality versus non-locality}
\label{sec:loc}
The discussion up to and including Theorem 1 proves the existence of efficiently universal gate sets in $SU(d)$ such that unitaries with $\Omega(\log{1/\epsilon})$ complexity are dense. One point that we have not considered so far is the issue of locality. For concreteness, consider a system of $N$ physical qubits on a ring, with Hilbert space dimension $d=2^N$, and suppose that the only experimentally accessible gates are those acting on at most two qubits at once. The space of all exact, all-to-all, two-qubit gates is described by $K = \mathcal{O}(N^2)$ continuous parameters. If one additionally imposes spatial locality, for example, allowing only two-qubit gates that act on nearest neighbour qubits, the number of continuous parameters shrinks to $K = \mathcal{O}(N)$.

A natural question is whether our results on the gate complexity can be formulated for sets of computationally universal gates $\mathcal{A}$ that are local in the above sense. This indeed turns out to be achievable, though the construction we propose is again somewhat indirect (and certainly not unique). Let us take ``local'' here to mean two-qubit gates acting on nearest neighbours, which allows for $K = N$ distinct gate configurations (note that our arguments extend straightforwardly to all-to-all gate sets). The first step is to choose a pair of rotations that are algebraic and densely generate $SU(2)$, for example $\{e^{i\pi \alpha Z}, e^{i\pi \alpha Y} \}$ with $\cos \pi \alpha= 1/3$. Then the embedding Eq. \eqref{eq:embedding} can be used to construct a gate set $\mathcal{A}$ with $6K$ elements that densely generates each nearest neighbour two-qubit copy of $SU(4)$. Since the full set of nearest-neighbour two-qubit gates is exactly universal~\cite{Piddock} in the full group $G = SU(2^N)$, the group $\langle \mathcal{A} \rangle$ must be dense in $G$. Moreover, each matrix in $\langle \mathcal{A} \rangle$ is algebraic by construction. It follows that the gate set $\mathcal{A}$ satisfies the hypotheses of Theorem \ref{thm:thm1}.

The upshot is that it is possible to construct local, efficiently universal gate sets such that unitaries with $\Omega(\log{1/\epsilon})$ complexity are dense in $G$. This is true whether one considers spatially local or all-to-all few-qubit gate sets. However, one shortcoming of our construction is that the geometry of the Cayley graph of $\langle \mathcal{A} \rangle$, which has been argued to capture important features of the quantum complexity~\cite{susskind2018lectures,Lin2019}, is not especially obvious.

If one instead relaxes the requirement of locality and asks only that the elementary gates be expressible as finite products of local operations, then a more explicit characterization of the Cayley graph becomes possible. Specifically, applying the Breuillard-Gelander theorem\cite{Breuillard,Breuillard2008} to the dense subgroup $\langle \mathcal{A} \rangle < G$ yields a pair of gates $A' = \{g_1,g_2\}$ that freely generate a dense subgroup of $SU(2^N)$, are expressible as finite products of local gates, and satisfy the hypotheses of Theorem \ref{thm:thm1}. In particular, the Cayley graph of $\langle \mathcal{A}' \rangle$ is a rooted Bethe lattice with coordination number $z=4$ (see Fig. \ref{Fig1} for a depiction), belying the physical intuition~\cite{susskind2018lectures} that the coordination number of the Cayley graph of a computationally universal gate set on $N$ qubits must scale with $N$.

\section{Complexity geometry}

\subsection{Riemannian complexity geometry}
We now turn to Nielsen's notion of complexity geometry\cite{nielsen2005geometric} for continuous sets of elementary gates $\mathcal{A}$, which applies ideas from the theory of Hamiltonian control to the problem of constructing arbitrary unitary operators through the repeated action of few-qubit gates. While the control theoretical formulation of quantum complexity allows for fairly arbitrary cost functions in principle, it is simplest to focus on cost functions that derive from an underlying Riemannian metric\cite{nielsen2005geometric,secondlaw}.

For concreteness let us again consider quantum algorithms on $N$ qubits. Then the Hilbert space dimension is $d=2^N$, and the goal is to describe the complexity geometry of the group $G=SU(d)$. We shall refer to elements of the Lie algebra $\frak{g}$ (viewed as the tangent space at the identity) in terms of Hermitian matrices, i.e. by writing $iH$, with $H$ Hermitian and traceless. Following Nielsen, let $\mathcal{P}$ denote a projector onto $k$-local Hermitian matrices with $k \leq 2$ and $\mathcal{Q}$ a projector onto those with $k>2$. Further define the ``penalty factor'' $q \geq 1$ and introduce an inner product
\begin{equation}
\label{eq:NielsenMetric}
\langle H_1, H_2 \rangle = \frac{\mathrm{Tr}(H_1 \mathcal{P}H_2) +  q\mathrm{Tr}(H_1 \mathcal{Q}H_2)}{2^N}
\end{equation}
on $\frak{g}$. By right translation, this can be extended to a Riemannian metric on all of $G$, and when $q=1$, recovers (up to a constant factor) the bi-invariant metric defined by the Killing form.

Using this inner product, we define the complexity distance of two unitary operators $U,V$ to be
\begin{equation}
C(U,V) = \inf_{\gamma \, \mathrm{a.c.}} \int_0^1 dt \, \langle H, H \rangle^{1/2},
\end{equation}
where the infimum is over all absolutely continuous paths $\gamma: [0,1] \to G$ such that
\begin{equation}
\label{eq:SE}
\frac{d}{dt} \gamma = -iH(t) \gamma, \quad \gamma(0) = U, \, \gamma(1) = V.
\end{equation}
For finite $q$, the complexity distance $C(U,V)$ is simply the geodesic distance from $U$ to $V$ with respect to the Riemannian metric Eq. \eqref{eq:NielsenMetric}. However, if one takes $q=\infty$ (the most physical choice from the viewpoint of locality), the inner product $\langle . \, , . \rangle$ ceases to be invertible.

In this limit, Nielsen's metric is most naturally interpreted as defining a sub-Riemannian geometry~\cite{Khaneja,nielsen2005geometric,subRiemRef}, for which only tangent directions consisting of $k$-body operators with $k\leq 2$ are allowed. This observation allows one to deduce several generic features of the $q=\infty$ limit using standard results on sub-Riemannian geometry, which do not seem to have been applied previously to the quantum complexity (although some of the resulting concepts were discussed for the special case of a single qubit~\cite{singlequbit}).

\subsection{Sub-Riemannian complexity geometry}
We first introduce some key definitions~\cite{ledonne2010lecture,Gromov1996}. Let $M$ be a real $n$-manifold and $\Delta \subset TM$ a smooth sub-bundle of the tangent bundle of $M$. The distribution $\Delta$ is called \emph{bracket generating} if any tangent vector $v \in T_{p}M$ can be expressed as a linear combination of vectors
$X_1, \, [X_2,X_3], \, [X_4,[X_5,X_6]], \ldots$ with $X_j \in \Delta$.  A \emph{sub-Riemannian} manifold consists of a triple $(M,\Delta,g)$ where $\Delta$ is a bracket-generating distribution and $g$ is a Riemannian metric on $M$. A curve $\gamma: [0,1] \to M$ is called \emph{horizontal} if $\dot{\gamma}(t) \in \Delta_{\gamma(t)}$ wherever $\dot{\gamma}$ is defined.

For Nielsen's geometry with $q=\infty$, we take $M=G$ and define $\Delta$ to be the space of $k$-body operators in $\frak{g}$ with $k\leq 2$, extended to all of $G$ by right translation. The metric $g$ on $G$ can be taken to be the standard bi-invariant Riemannian metric
\begin{equation}
\label{eq:Riemannian}
\langle H_1, H_2 \rangle = \frac{\mathrm{Tr}(H_1 H_2)}{2^N},
\end{equation}
though more general choices are allowed. Then the triple $(G,\Delta,g\big{|}_\Delta)$ is a sub-Riemannian manifold. Given these definitions, the complexity distance of two unitary operators $U,V$ is defined as the length of the shortest horizontal curve connecting $U$ to $V$, to wit
\begin{equation}
C(U,V) = \inf_{\substack{\gamma \mathrm{\,a.c.} \\\dot{\gamma}\subset \Delta}} \int_0^1 \, \langle H , H \rangle^{1/2},
\end{equation}
with $\gamma$ satisfying Eq. \eqref{eq:SE}. Letting $d(U,V)$ denote the geodesic distance with respect to the standard ($q=1$) metric, Eq. \eqref{eq:Riemannian}, it is clear that the complexity distance is bounded below by the usual geodesic distance, i.e.
\begin{align}
\label{eq:lowerbound}
C(U,V) \geq d(U,V).
\end{align}
The metric defined by $C(U,V)$ is usually called a Carnot-Carath{\'e}odory distance~\cite{Gromov1996} (though this seems to be something of a misnomer, given that the distributions appearing in Carath{\'e}odory's formulation of the second law of thermodynamics are integrable~\cite{frankel2011geometry}).

When the distribution $\Delta$ is bracket generating, the following result holds:

\begin{theorem} \emph{(Chow\cite{Gromov1996})} Let $M$ be a connected manifold and $\Delta$ a bracket-generating distribution. Then any two points in $M$ can be joined by a piecewise smooth curve that is horizontal.
\end{theorem}
In the context of Nielsen's complexity geometry, this means that for any $U,V \in G$, there exists a piecewise smooth curve from $U$ to $V$ such that the instantaneous Hamiltonian $H(t)$ comprises purely $k$-body terms with $k\leq 2$. In particular, the complexity distance $C(U,V)$ is always finite. However, these curves are expected to look much ``rougher'' than geodesics in the Riemannian metric. This intuition can be made precise as follows.

Let us first define the flag of sub-bundles
\begin{equation}
\Delta \subset \Delta^{[2]} \subset \Delta^{[3]} \ldots \subset \Delta^{[s]} = TG
\end{equation}
inductively by allowing successive commutators. Namely, we set $\Delta^{[1]} = \Delta$ and define
\begin{equation}
\Delta^{[k+1]} = \mathrm{span}\{\Delta^{[k]}, [X, Y] : X \in \Delta, Y \in \Delta^{[k]}\}, \quad k > 1.
\end{equation}
Thus, for example,
\begin{equation}
\Delta^{[2]} = \mathrm{span}\{X_1, [X_2, X_3] : X_1,X_2,X_3 \in \Delta\}.
\end{equation}

Now let $X_1,X_2,\ldots,X_n$ be a frame of sections of $TG$ and define integers $m_k$ such that $X_1,X_2,\ldots,X_{m_k}$ is a frame of sections for $\Delta^{[k]}$ (thus $m_k$ is the number of linearly independent vectors in $\Delta^{[k]}$ at a given point). The \emph{degree} $d_j$ of the vector field $X_j$ quantifies its commutator depth, and is defined so that
\begin{equation}
X_j \in \Delta^{[d_j]} \backslash \Delta^{[d_j-1]}.
\end{equation}
with $\Delta^{[0]} := \{0\}$. In the complexity language, vector fields with $d_j = 1$ correspond to ``easy'' directions in the space of Hermitian operators while vectors with $d_j >1$ correspond to ``hard'' directions. One of the basic results of sub-Riemannian geometry relates this structure of ``easy'' and ``hard'' directions to the local behaviour of the complexity metric.

To formulate this result, let us first define the flow $\mathrm{exp}_U(X)$ that takes $U$ to the point $\gamma(1)$ on the integral curve
\begin{equation}
\frac{d}{dt} \gamma(t) = X_{\gamma(t)}, \quad \gamma(0) = U.
\end{equation}
This defines an exponential map $\mathrm{exp}_U : \mathbb{R}^n \to G$, of the form
\begin{equation}
(t_1,t_2,\ldots,t_n) \mapsto \mathrm{exp}_U(t_1X_1+t_2X_2+\ldots + t_n X_n).
\end{equation}
We define the ``box'' $\mathrm{Box}(r)$ in $\mathbb{R}^n$ as
\begin{equation}
\mathrm{Box}(r) = \{(t_1,t_2,\ldots,t_n) \in \mathbb{R} : |t_j| \leq r^{d_j}\}.
\end{equation}
These anisotropic boxes in $\mathbb{R}^n$ capture the geometry of balls $B_U(r) = \{V \in G : C(U,V) < r\}$ in the complexity metric. In particular, the following result holds:

\begin{theorem} \emph{(the ``ball-box theorem''\cite{Gromov1996})} Balls in the complexity metric are uniformly equivalent to images of boxes under the exponential map. Specifically, there are strictly positive continuous functions $A, r: G \to \mathbb{R}^+$ with $A>1$ such that 
\begin{equation}
\label{eq:boxball}
\mathrm{exp}_U \mathrm{Box}(A^{-1}r) \subset B_{U}(r) \subset \mathrm{exp}_U \mathrm{Box}(Ar).
\end{equation}
\end{theorem}
This result has two important consequences for sub-Riemannian complexity geometry. First, it implies this geometry is fractal, in the sense that the Hausdorff dimension of the metric space $(G,C)$ is equal to~\cite{Gromov1996} $n_H = \sum_{j=1}^n d_j > n$, and therefore exceeds its real dimension $n=4^N-1$. This fact was recently noted in the literature without proof~\onlinecite{susskind2020black}. It follows intuitively from Eq. \eqref{eq:boxball}, which suggests that the volume of a complexity ball scales anomalously with its radius $r$, as $\sim r^{\sum_{j=1}^n d_j}$.

Second, this result suggests that moving in a generic direction from a unitary $U$ to another unitary $V$ at a Riemannian distance $d(U,V) = \delta$, the complexity distance behaves as $C(U,V) \sim \delta^{1/s}$, where $s=\max_j d_j >1$. More formally, the identity map relating the metric spaces $\mathrm{id}: (G,d) \to (G,C)$ is known\cite{Gromov1996,ledonne2010lecture} to be $\alpha$-H{\"o}lder continuous with H{\"o}lder exponent $\alpha = 1/s$, which implies the inequality
\begin{equation}
\label{eq:Holder}
C(U,V) \leq M_U \delta^{1/s}
\end{equation}
for some constant $M_U>0$ and $\delta$ sufficiently small. This indicates that the complexity distance is continuous but possibly not differentiable as $V \to U$. A precise statement is as follows:

\begin{prop} 
\label{prop:nielsen}
The $q=\infty$ complexity distance $C(U,V)$ is continuous as $V \to U$ but does not have one-sided directional derivatives as $V \to U$ in generic directions.
\end{prop}

\emph{Proof}. First introduce ``privileged coordinates''\cite{Bellaiche1996} $y_1,y_2,\ldots,y_n$ about the point $U$, which satisfy
\begin{align}
a_U (|y_1|^{1/d_1}+|y_2|^{1/d_2} + \ldots + |y_{n}|^{1/d_n}) \leq C(U,V)
\label{eq:ineq}
\end{align}
for some constant $a_U>0$. Here, $y_1,\,y_2,\ldots,y_n$ denote the privileged coordinates of $V$, while $U$ has privileged coordinates $\mathbf{y}=0$ by construction. Choosing $V$ to approach $U$ in a generic direction $\mathbf{v}$, i.e. defining $V(t)$ by its privileged coordinates $y_j = v_jt$ with $|v_j|>0$ and sufficiently small $t$, we have
\begin{equation}
\frac{C(U,V(t))}{t}  > a_U |v_n| t^{1/s-1} \to \infty, \quad t \to 0^+.
\end{equation}
We deduce that approaching from generic directions in the tangent space at $U$, the one-sided directional derivative $\lim_{t\to0^+} \frac{C(U,V(t))}{t}$ does not exist. Meanwhile the H{\"o}lder condition Eq. \eqref{eq:Holder} implies that $C(U,V) \to 0$ as $V \to U$. $\qed$

At this point, several comments are in order. First, the results described above can be adapted to sub-Finsler metrics\cite{ledonne2010lecture}, and in this sense capture the most general case of Nielsen's complexity geometry for bracket-generating distributions $\Delta$. Second, the lack of one-sided directional derivatives of the sub-Riemannian complexity distance $C(U,V)$ as $V \to U$ distinguishes it from the Riemannian geodesic distance $d(U,V)$. While neither function is differentiable at $V = U$, the Riemannian distance admits one-sided directional derivatives as $V \to U$ along smooth curves.

We now briefly discuss the question of local versus non-local gate sets in the context of the sub-Riemannian complexity distance. As noted above, the local geometry of complexity balls is entirely determined by the structure of ``easy'' and ``hard'' vector fields encoded by the sequence of integers $(d_1,d_2,\ldots,d_n)$. Suppose now that in Nielsen's definition we impose spatial locality rather than two-locality, i.e. demand that the ``easy'' directions in $\frak{g}$ comprise only spatially local two-qubit operators. As discussed in Section \ref{sec:loc}, this modifies the dimension of the vector space $K = \mathrm{dim}(\Delta)$ of easy directions from $K = \mathcal{O}(N^2)$ for all-to-all, two-local gates to $K = \mathcal{O}(N)$ for spatially local two-qubit gates. This in turn leads to a redefinition of the integers $(d_1,d_2,\ldots,d_n)$ and a corresponding increase in the Hausdorff dimension $n_H$, but since both distributions are bracket generating~\cite{Piddock} there are no other substantive changes.

\subsection{Approaching the $q=\infty$ limit}
For large but finite $q$, Nielsen's metric Eq. \eqref{eq:NielsenMetric} is Riemannian. This means that the complexity distance function $C(U,V)$ admits one-sided directional derivatives along smooth curves through $U$ (though it is still not differentiable at $V=U$). Moreover, the complexity geometry is no longer truly fractal because the Riemannian exponential map is locally well-behaved~\cite{frankel2011geometry}, raising the question of how far the singular features of the $q=\infty$ limit are inherited by the complexity geometry with $q<\infty$. This point was previously discussed for the special case of a single qubit~\cite{singlequbit}. We now generalize these considerations to the complexity geometry of $N$ qubits; our discussion will be less rigorous than in preceding sections.

As noted previously~\cite{susskind2018lectures,singlequbit,susskind2020black}, the relevant geometrical idea is the ``cut locus'' $\Gamma_U$ of a given $U \in G$, which is the set of points $V \in G$ for which there exists more than one length-minimizing geodesic connecting $U$ and $V$. For large $q$, we expect that the nearest points $V^* \in \Gamma_U$ to $U$ have the property that the sub-Riemannian geodesic connecting $U$ to $V^*$ is equal in length to the Riemannian geodesic connecting $U$ to $V^*$. Writing $d(U,V^*) = \delta$, we therefore expect
\begin{equation}
C(U,V^*) \sim q^{1/2}\delta \sim \delta^{1/s}
\end{equation}
in a generic direction, where $s$ was defined above.

Thus the distance $\delta$ to the nearest cut point $V \neq U$ such that $C(U,V)$ is not smooth scales with $q$ as
\begin{equation}
\delta \sim q^{-s/(2(s-1))}.
\end{equation}
For a single qubit with $s=2$, this recovers a previous result~\cite{singlequbit}. Notice that as $q \to \infty$, the cut locus $\Gamma_U$ becomes arbitrarily close to $U$ in almost all directions. This is one way to understand the singular nature of the limit $q=\infty$.

\subsection{Complexity distance versus gate complexity}
\label{sec:obstacle}

Finally, one can ask how far the continuous complexity distance emerges naturally as a continuum limit of the discrete gate complexity. The existence of such a limit would be strong evidence in favour of a universal notion of quantum complexity. 

Remarkably, an equivalence between discrete gate complexity (i.e. a word metric) and continuous complexity distance (i.e. a Carnot-Carath{\'e}odory metric) \emph{does} exist for nilpotent Lie groups, for example, the Heisenberg group. Roughly speaking, this arises due to an equivalence between the volume of Cayley graphs at radius $r$ and the volume of their limiting sub-Riemannian balls at radius $r$, both of which scale with the same anomalous Hausdorff dimension, e.g. as $\sim r^4$ for the Heisenberg group~\cite{Gromov1981,pansu1983croissance}.

However, for unitary groups any such local equivalence between discrete and continuous complexity geometries breaks down. This is easily seen by comparing the volume growth of the Cayley graph on two free generators at radius $r$, which scales exponentially as $\sim 3^r$, with the volume of a complexity ball, which scales polynomially as $\sim r^{n_H}$. The existence of exponentially growing Cayley graphs in unitary Lie groups, compared to nilpotent ones, reflects the basic distinction between semisimple and solvable Lie algebras, and can be viewed as a consequence of the Tits alternative in group theory~\cite{Tits,Gromov1981}. This appears to rule out a correspondence between discrete and continuous notions of quantum complexity down to the smallest scales of the unitary group $G$ in question.

\section{Discussion}
We have proposed a unifying perspective on the various notions of quantum complexity and shown how they can be distinguished by their smoothness properties, or lack thereof. Our results clarify how the gate complexity tends to a nowhere continuous function of its argument as $\epsilon \to 0$, and make precise the sense in which Nielsen's complexity geometry becomes fractal as $q \to \infty$.

At the time of writing, it is still not clear how far there exists a ``unique'', or universal, definition of quantum complexity, nor indeed whether such a definition can be expressed in terms of the complexity measures studied in our paper. One possible route to tackling this question would be making sense of how the different notions of complexity depicted in Fig. \ref{Fig1} emerge as limits of one another. The breadth of mathematical ideas involved in their formulation, together with the group-theoretical obstruction identified above, suggests that this task will be far from trivial. Meanwhile, the idea that there might exist such a universal notion of quantum complexity~\cite{ComplexityEqualsAction} motivates further study of complexity measures in realistic many-body quantum systems.

\emph{Acknowledgments.} We thank V. Khemani, S. Parameswaran, R. Nandkishore, R. Moessner and especially A. Brown for helpful discussions. We are grateful to A. Harrow for comments on the manuscript. This work was supported with funding from the Defense Advanced Research Projects Agency (DARPA) via the DRINQS program.
\bibliography{bibl.bib}
\end{document}